\newcommand{\bee}{\begin{equation}}
\newcommand{\ene}{\end{equation}}
\newcommand{\beea}{\begin{eqnarray}}
\newcommand{\enea}{\end{eqnarray}}

\documentclass[aps,preprint,pre,showpacs,superscriptaddress]{revtex4}
\usepackage{amsmath} 
\usepackage{graphicx}
\usepackage{dcolumn}
\usepackage{bm}
\begin{document}
\title{Resonant Electron-Plasmon Interactions in Drifting Electron Gas}
\author{M. Akbari-Moghanjoughi}
\affiliation{Faculty of Sciences, Department of Physics, Azarbaijan Shahid Madani University, 51745-406 Tabriz, Iran}

\begin{abstract}
In this paper we investigate the resonant electron-plasmon interactions in a drifting electron gas of arbitrary degeneracy. The kinetic-corrected quantum hydrodynamic model is transformed into the effective Schr\"{o}dinger-Poisson model and driven coupled pseudoforce system is obtained via the separation of variables from the appropriately linearized system. It is remarked that in the low phase-speed kinetic regime the characteristic particle-like plasmon branch is significantly affected by the correction factor which is a function of the electron number density and temperature. It is shown that electron current density of drifting electron gas sharply peaks at two distinct drift wavenumbers for given value of electron density, temperature, plasmon energy and damping parameter. The Fano-resonance of current density profile confirms the electron-plasmon resonant interaction in the presence of underlying interference effect. The electron current density also shows a very sharp peak at resonant plasmon energy. Moreover, an extension to multistream model is presented and the total current density of drifting electron gas in the presence of resonant electron-plasmon interactions is obtained. We further investigate the kinetic correction effect on matter-wave energy dispersion of the electron gas. It is also found that increase in the electron number density leads to increase in effective mass and consequently decrease in electron mobility, whereas, increase in the electron temperature has the converse effect. The kinetic correction significantly lowers the plasmon conduction band minimum. Current model may be further elaborated to investigate the electron beam-plasma interactions.
\end{abstract}

\date{\today}
\email[Corresponding author: ]{massoud2002@yahoo.com}
\maketitle

\section{Introduction}

Plasma oscillations play fundamental role in important physical phenomena associated with collective interaction of charged species. Due to complex nature of electromagnetic interactions between different charges in multispecies plasmas, a wide spectrum of interesting linear and nonlinear phenomena such as a wide range of instabilities \cite{chen}, wave-particle interactions \cite{krall}, solitons, double layer, shock waves \cite{drazin}, etc. can take place in these environments. New developments in quantum plasma theories such a kinetic and hydrodynamic models, has been used in order to study various collective quantum effects \cite{scripta,manfredi}. The wave-kinetic approach to the Schr\"{o}dinger-Newton model, as a generalized Schr\"{o}dinger-Poisson model, has been extensively studied in Ref. \cite{mend}. In the later model, the extension to the relativistic particles has also been discussed as a generalized Klein-Gordon equation. In plasmas the electron fluid is the main inertialess ingredient with dielectric response to electromagnetic waves and dynamic structure factor \cite{ichimaru1,sturm} leading to many important phenomena, such as, Thomson, Compton \cite{jung3}, stimulated Raman and Brillouin \cite{son,roz} scattering which are used as efficient plasma diagnostic methods. Electrons are also responsible for other important physical properties in plasmas, such as, charge shielding \cite{zhao,mold0,hong,ydj1,ydj2,yoon}, ionic bound states \cite{yu,sahoo,kar}, electrical and energy transport phenomena, heat capacity and many others. Dielectric response of electron gas in solid state plasmas plays the main role in optical properties of metals \cite{kit,ash}, semiconductors \cite{hu1,seeg}, nanometallic structures, low dimensional systems and liquid crystals \cite{lucio,mark,haug,gardner,man1,maier,yofee}. However, in order to understand many of these physical properties, one has to obtain detailed information on collective aspects of electron fluid. In hydrodynamic plasma theories the electron dynamic is simply coupled to electromagnetic fields through Maxwell equations. Therefore, hydrodynamic and magnetohydrodynamic theories constitute a cost effective and analytic means to investigate a wide range of plasma phenomena in a straightforward manner \cite{haas1}. These theories, if properly formulated, are also believed to be able to capture many collective aspect such as collisionless or collisional damping, previously known to be purely kinetic effects \cite{mannew}.

Due to growing necessity for developments in miniaturized low-dimensional semiconductor and nanostructured devices for futuristic electronic purposes, a renewed effort has been devoted to discover physical properties of dense quantum electron gas, over the past few years. Although the theories of quantum electron gas dielectric response has been fully developed more than half a century ago \cite{madelung,bohm,bohm1,bohm2,pines,levine,klimontovich}, many interesting properties of quantum electron fluid is yet to be discovered. Many improved and enhanced quantum kinetic and quantum hydrodynamic theories have emerged over the recent years. Application of these theories predicts a very insightful future for the science of quantum plasmas and have helped to discover a broad range of new interesting collective phenomena of quantum electron gas, unknown for many years \cite{se,sten,ses,brod1,mark1,man3,mold2,sm,hurst,fhaas,kim}. Application of the Schr\"{o}dinger-Poisson system and the corresponding coupled pseudoforce model has recently been used to study various features of collective excitations in ideal electron gas \cite{akbquant,akbfano,akbdual}. In current research, starting from kinetic corrected quantum hydrodynamic model of an electron gas in one dimension, a system of coupled pseudoforce system is derived for arbitrary degenerate electron gas and the resonant interaction of the drifting electrons with plasmon excitations (quasiparticles) is studied.

\section{The Quantum Hydrodynamic Model}

Dynamic properties of a free electron gas can be studied through the hydrodynamic model to a great extent. Quantum hydrodynamic model may be used to study the dynamic properties of dense electron gas with arbitrary degree of degeneracy, as well. A closed set of conventional quantum hydrodynamic equations for isothermal electron gas, reads
\begin{subequations}\label{gs}
\begin{align}
&\frac{{\partial n}}{{\partial t}} + \frac{{\partial nv}}{{\partial x}} = 0,\\
&\frac{{\partial v}}{{\partial t}} + v\frac{{\partial v}}{{\partial x}} = \frac{e}{m}\frac{{\partial \phi }}{{\partial x}} - \frac{1}{m}\frac{{\partial \mu }}{{\partial x}} + {\rm{ }}\frac{{{\gamma\hbar ^2}}}{{2{m^2}}}\frac{\partial }{{\partial x}}\left( {\frac{1}{{\sqrt n }}\frac{{{\partial ^2}\sqrt n }}{{\partial {x^2}}}} \right),\\
&\frac{{{\partial ^2}\phi }}{{\partial {x^2}}} = 4\pi e\left( {n - {n_0}} \right),
\end{align}
\end{subequations}
where the dependent variables, $n$, $v$, $\mu$ and $\phi$ refer to the number density, fluid velocity, chemical potential and electrostatic potential, respectively. The last term in the momentum equation is the well-known Bohm potential which leads to quantum diffraction effects. The prefactor $\gamma$ is the kinetic correction parameter applying to the Bohm potential, which has been the subject of intense debate over the recent years \cite{bonitz1,sea1,bonitz2,sea2,bonitz3,akbhd,bonitz0}. It is remarked that, in the fully degenerate regime, the kinetic correction approaches the limiting value, $\gamma=1/9$ \cite{akbhd,moldabekov}. For our purpose, we use an ideal gas without account for the electron exchange effect and assume that their interactions are only via the electrostatic potential. The parametric equation of state (EOS) for slow isothermal compression of electrons with arbitrary degree of degeneracy can be given in terms of the Fermi integrals
\begin{subequations}\label{eosr}
\begin{align}
&n(\eta ,T) = \frac{{{2^{7/2}}\pi {m^{3/2}}}}{{{h^3}}}{F_{1/2}}(\eta) =  - \frac{{{2^{5/2}}{{(\pi m{k_B}T)}^{3/2}}}}{{{h^3}}}{\rm{L}}{{\rm{i}}_{3/2}}[ - \exp (\eta )],\\
&P(\eta ,T) = \frac{{{2^{9/2}}\pi {m^{3/2}}}}{{3{h^3}}}{F_{3/2}}(\eta) =  - \frac{{{2^{5/2}}{{(\pi m{k_B}T)}^{3/2}}({k_B}T)}}{{{h^3}}}{\rm{L}}{{\rm{i}}_{5/2}}[ - \exp (\eta )],
\end{align}
\end{subequations}
where $\eta=\beta\mu$ with $\beta=1/k_B T$ and $F_k$ is the Fermi integral of order $k$
\begin{equation}\label{f}
{F_k}(\eta ) = \int_0^\infty  {\frac{{{x^k}}}{{\exp (x - \eta ) + 1}}} dx.
\end{equation}
The function ${\rm{Li}}_{k}$ is the polylog function defined in terms of Fermi integrals
\begin{equation}\label{pl}
{F_k}(\eta ) =  - \Gamma (k + 1){\rm{L}}{{\rm{i}}_{k + 1}}[ - \exp (\eta )],
\end{equation}
in which $\Gamma$ is the gamma function. Note that the isothermal EOS is not the only choice, whereas, for fast processes the adiabatic EOS \cite{elak} should be applicable. However, for simplicity, we have chosen to use the isothermal electron EOS in current model which is based on Thomas-Fermi model \cite{dya1,dya2}. The kinetic correction has been studied in many recent literature \cite{michta,akbhd,sm,mold1,moldabekov,haas2016} and for electron gas with arbitrary degeneracy at finite temperature may be written as \cite{haas2016}
\begin{equation}\label{xi}
\gamma  = \frac{{{\rm{L}}{{\rm{i}}_{3/2}}\left[ {-\exp (\beta {\mu _0})} \right]{\rm{L}}{{\rm{i}}_{ - 1/2}}\left[ {-\exp (\beta {\mu _0})} \right]}}{3{{\rm{L}}{{\rm{i}}_{ 1/2}}{{\left[ {-\exp (\beta {\mu _0})} \right]}^2}}},
\end{equation}
where, $\mu_0$ is the equilibrium chemical potential. In the limit $\mu_0=0$, we have $\gamma=1/3$. The quantum hydrodynamic model (\ref{gs}) may be transformed \cite{manfredi} to the following Schr\"{o}dinger-Poisson system using the Madelung transformations \cite{madelung}
\begin{subequations}\label{sp}
\begin{align}
&i\hbar\sqrt{\gamma}\frac{{\partial \cal N }}{{\partial t}} =  - \frac{\gamma{{\hbar ^2}}}{{2m}}\frac{{\partial ^2 {{\cal N}}}}{{\partial {x^2}}} - e\phi{\cal N} + \mu(n,T){\cal N},\\
&\frac{{\partial ^2 {\phi}}}{{\partial {x^2}}} = 4\pi e (|{\cal N}|^2 - n_0),
\end{align}
\end{subequations}
where ${\cal N} =\sqrt{n(x,t)}\exp[iS(x,t)/\hbar\sqrt{\gamma}]$ is the state-function characterizing spatiotemporal evolution of the collective electron excitations with ${\cal N}{\cal N^*}=n(x,t)$ being the number density and $v(x,t)=(1/m)\partial S(x,t)/\partial x$ the electron fluid velocity. Also, $n_0$ represents the equilibrium electron gas number density. We will use the coupled system, (\ref{sp}), to study the interaction of an electron drift with collective excitations we call quasiparticles, hereafter. Let us assume the electrons drifting at constant fluid velocity, $v$. Using the form ${\cal N} =\sqrt{n(x,t)}\exp[iS(x,t)/\hbar\sqrt{\gamma}]$ in (\ref{sp}) and by separation of real and imaginary parts, one arrives at the following relations
\begin{subequations}\label{sph}
\begin{align}
&m\frac{{\partial n(x,t)}}{{\partial t}}+\frac{{\partial n(x,t)}}{{\partial x}}\frac{{\partial S(x,t)}}{{\partial x}}+n(x,t)\frac{{{\partial ^2}S(x,t)}}{{\partial {x^2}}} = 0,\\
&\frac{{{\partial ^2}S(x,t)}}{{\partial t\partial x}} + \frac{1}{m}\frac{{\partial S(x,t)}}{{\partial x}}\frac{{{\partial ^2}S(x,t)}}{{\partial {x^2}}} = \frac{{e\partial \phi (x,t)}}{{\partial x}} - \frac{{\partial \mu (x)}}{{\partial x}} + \frac{{\partial B(x,t)}}{{\partial x}},\\
&B(x,t) = \frac{{\gamma {\hbar ^2}}}{{8m{n^2}(x,t)}}\left\{ {2n(x,t)\frac{{{\partial ^2}n(x,t)}}{{\partial {x^2}}} - {{\left[ {\frac{{\partial n(x,t)}}{{\partial x}}} \right]}^2}} \right\},
\end{align}
\end{subequations}
which are in full agreement with the continuity and momentum hydrodynamic equations in (\ref{gs}) by the definition, $v(x,t)=(1/m)\partial S(x,t)/\partial x$.

The value of the electron chemical potential may vary depending on the degeneracy degree and the temperature in a wide range. However, in a classical regime, it can have negative values up to zero electronvolts, whereas, in quantum regime the value will be typically positive few electronvolts, such as, for metals . The Schr\"{o}dinger-Poisson system (\ref{sp}), thus, can model a wide range of phenomenon with electron excitations ranging from classical regime up to the fully degenerate quantum case. For our purpose we have, $S(x,t)=p x+f(t)$ in which $p=mv$ is the electron fluid momentum and $f(t)$ is an arbitrary function of time. Let us now consider separable solutions of the form, ${\cal N}(x,t)={\psi}(t){\psi}(x)\exp(ipx/\hbar\sqrt{\gamma})$ with ${\psi}(x)=\sqrt{n(x)}$. Separation of spatial and temporal parts in first equation of (\ref{sp}), leads to
\begin{subequations}\label{ssp}
\begin{align}
&\frac{{\gamma {\hbar ^2}}}{{2m}}\frac{{{d^2}}}{{d{x^2}}}\psi (x){e^{\frac{{ipx}}{{\hbar \sqrt \gamma  }}}} + \left[ {e\phi (x) + \epsilon - \mu (n,T)} \right]\psi (x){e^{\frac{{ipx}}{{\hbar \sqrt \gamma  }}}} = 0,\\
&\frac{{d^2 {\phi(x)}}}{{d{x^2}}} = 4\pi e (|{\psi}(x)|^2 - n_0),\\
&i\hbar\sqrt{\gamma}\frac{{d{\psi}(t)}}{{dt}} = \epsilon {\psi}(t),
\end{align}
\end{subequations}
where $\epsilon$ is the energy eigenvalue of the quantum system. Here, we are interested in finding solutions for spatial variations in quantities such as density and electric potential. Therefore, we only have the first two coupled equations in (\ref{ssp}) in order to solve. The appropriately linearized system for small perturbations, assuming $\psi\simeq \psi_0+\psi_1$, $\phi\simeq 0+\phi_1$ and $p=0+p_1$, appears as the following coupled pseudoforce system
\begin{subequations}\label{lssp}
\begin{align}
&\frac{{\gamma {d^2}\Psi (x)}}{{d{x^2}}} + \Phi (x) + E\Psi (x) = {\cal E}\exp \left( {\frac{{i{k_d}x}}{{\sqrt \gamma  }}} \right),\\
&\frac{{{d^2}\Phi (x)}}{{d{x^2}}} - \Psi (x) = 0,
\end{align}
\end{subequations}
in which ${k_d} = p/\hbar {k_p}$ is the normalized drift wavenumber with $k_p=\sqrt{2m E_p}/\hbar$ being the plasmon length, $E_p=\hbar\omega_p$ the plasmon energy and $\omega_p=\sqrt{4\pi e^2n_0/m}$ the electron plasma frequency. We have also used the normalization scheme, $\Psi(x)={\psi}(x)/{\psi_0}$ where ${\psi_0}=\sqrt{n_0}$ is the unperturbed value of functional, ${\psi}(x)$ and $\Phi(x)=e\phi/E_p$ is normalized electrostatic energy. Also, ${E}=(\epsilon-\mu_0)/E_p$ and ${\cal E}=E_d-E=k_d^2-E$ with $E_d=p^2/2mE_p$ being the normalized drift energy. The spacial coordinate $x$ is normalized to the plasmon wavelength, $\lambda_p=2\pi/k_p$, with $k_p=\sqrt{2mE_p}/\hbar$ being the characteristic plasmon wavenumber. Moreover, the time is scaled with respect to the inverse plasmon frequency, $\omega_p=E_p/\hbar$. Note that the solution would be quasi-stationary, hence, the spacial distribution of electron density remains constant over time, whereas, the electron drift will be uniform.

\section{Resonant Plasmon Excitations}

We first consider the solution to the coupled psuedoforce system (\ref{lssp}) with real part of the driving force, for illustration purpose. The general solution to the homogenous system, reads
\begin{equation}\label{wf}
\left[ {\begin{array}{*{20}{c}}
{{\Phi _g}(x)}\\
{{\Psi _g}(x)}
\end{array}} \right] = \frac{\gamma }{\alpha }\left[ {\begin{array}{*{20}{c}}
{{\Psi _0} + k_2^2{\Phi _0}}&{ - \left( {{\Psi _0} + k_1^2{\Phi _0}} \right)}\\
{ - \left( {{\Phi _0}/\gamma  + k_1^2{\Psi _0}} \right)}&{{\Phi _0}/\gamma  + k_2^2{\Psi _0}}
\end{array}} \right]\left[ {\begin{array}{*{20}{c}}
{\cos ({k_1}x)}\\
{\cos ({k_2}x)}
\end{array}} \right],
\end{equation}
in which $\Phi_0$ and $\Psi_0$ characterize the initial values, assuming, $d\Psi(x)/dx|_{x=0}=d\Phi(x)/dx|_{x=0}=0$, for simplicity. The characteristic quasiparticle excitation wavenumbers $k_1$ and $k_2$ are
\begin{equation}\label{ks}
{k_1} = \sqrt {\frac{{E  - \alpha }}{{2\gamma }}},\hspace{3mm}{k_2} = \sqrt {\frac{{E  + \alpha }}{{2\gamma }}},\hspace{3mm}\alpha  = \sqrt {{E^2} - 4\gamma }.
\end{equation}
Note the complementarity relation, $k_1k_2=1$ between the dual lengths. The presence of dual lengthscale character of plasmon excitations has been previously reported by Ali etal. \cite{ali1,ali2}, using the quantum hydrodynamic analysis. It can be seen that the excitations satisfy the energy dispersion relation, $E=(1+\gamma k^4)/k^2$, with the minimum values at $(E_m,k_m)=(2\gamma^{1/2},\gamma^{-1/4})$. However, note that the energy dispersion relation obtained from the system (\ref{wf}) is different from the plasmon dispersion derived linearized quantum hydrodynamic model (\ref{gs}). In the inertial reference moving with drifting electrons, one has $v=0$ and one arrives at the following system of equations in the quantum hydrodynamic model
\begin{subequations}\label{gs1}
\begin{align}
&\frac{{\partial n}}{{\partial t}} = 0,\\
&g(t) = e\phi - \mu + \frac{{\gamma {\hbar ^2}}}{{2m\sqrt n }}\frac{{{\partial ^2}\sqrt n }}{{\partial {x^2}}},\\
&\frac{{{\partial ^2}\phi }}{{\partial {x^2}}} = 4\pi e\left( {n - {n_0}} \right),
\end{align}
\end{subequations}
in which $g(t)$ is an arbitrary function of time. Evidently the system (\ref{gs1}) does not give rise to the energy dispersion obtained by separation of variables and exact solution of the Schr\"{o}dinger-Poisson system (\ref{wf}). Therefore, current model of plasmon excitation should not be confused with the theory of electron plasma oscillations studied in conventional hydrodynamic framework. one of the main advantages of the Schr\"{o}dinger-Poisson system is that it provides an alternative approach to the electronic excitations as quasiparticles. The particular solution to the system (\ref{wf}) can be written in the following form
\begin{subequations}\label{psol}
\begin{align}
&{\Phi _p}(x) = \frac{{{\cal E}\left[ {\alpha \cos \left( {{k_d}x/\sqrt{\gamma}} \right) + \gamma \left( {k_d^2/\gamma - k_2^2} \right)\cos \left( {{k_1}x} \right) - \gamma \left( {k_d^2/\gamma - k_1^2} \right)\cos \left( {{k_2}x} \right)} \right]}}{{\alpha \gamma \left( {k_d^2/\gamma - k_1^2} \right)\left( {k_d^2/\gamma - k_2^2} \right)}},\\
&{\Psi _p}(x) = \frac{{{\cal E}\left[ {\left( {\alpha k_d^2/\gamma } \right)\cos \left( {{k_d}x/\sqrt \gamma  } \right) + \left( {1 - k_d^2k_1^2} \right)\cos \left( {{k_1}x} \right) - \left( {1 - k_d^2k_2^2} \right)\cos \left( {{k_2}x} \right)} \right]}}{{\alpha \gamma \left( {k_d^2/\gamma  - k_1^2} \right)\left( {k_d^2/\gamma  - k_2^2} \right)}}.
\end{align}
\end{subequations}
Note that, the kinetic correction in our normalization scheme, takes the following form
\begin{equation}\label{xin}
\gamma  = \frac{{{\rm{L}}{{\rm{i}}_{3/2}}\left[ { - \exp (\sigma /\theta )} \right]{\rm{L}}{{\rm{i}}_{ - 1/2}}\left[ { - \exp (\sigma /\theta )} \right]}}{3{{\rm{L}}{{\rm{i}}_{1/2}}{{\left[ { - \exp (\sigma /\theta )} \right]}^2}}},
\end{equation}
in which $\sigma=\mu_0/E_p$ and $\theta=T/T_p$ with $T_p=E_p/k_B$ being the characteristic plasmon temperature.

\begin{figure}[ptb]\label{Figure1}
\includegraphics[scale=0.7]{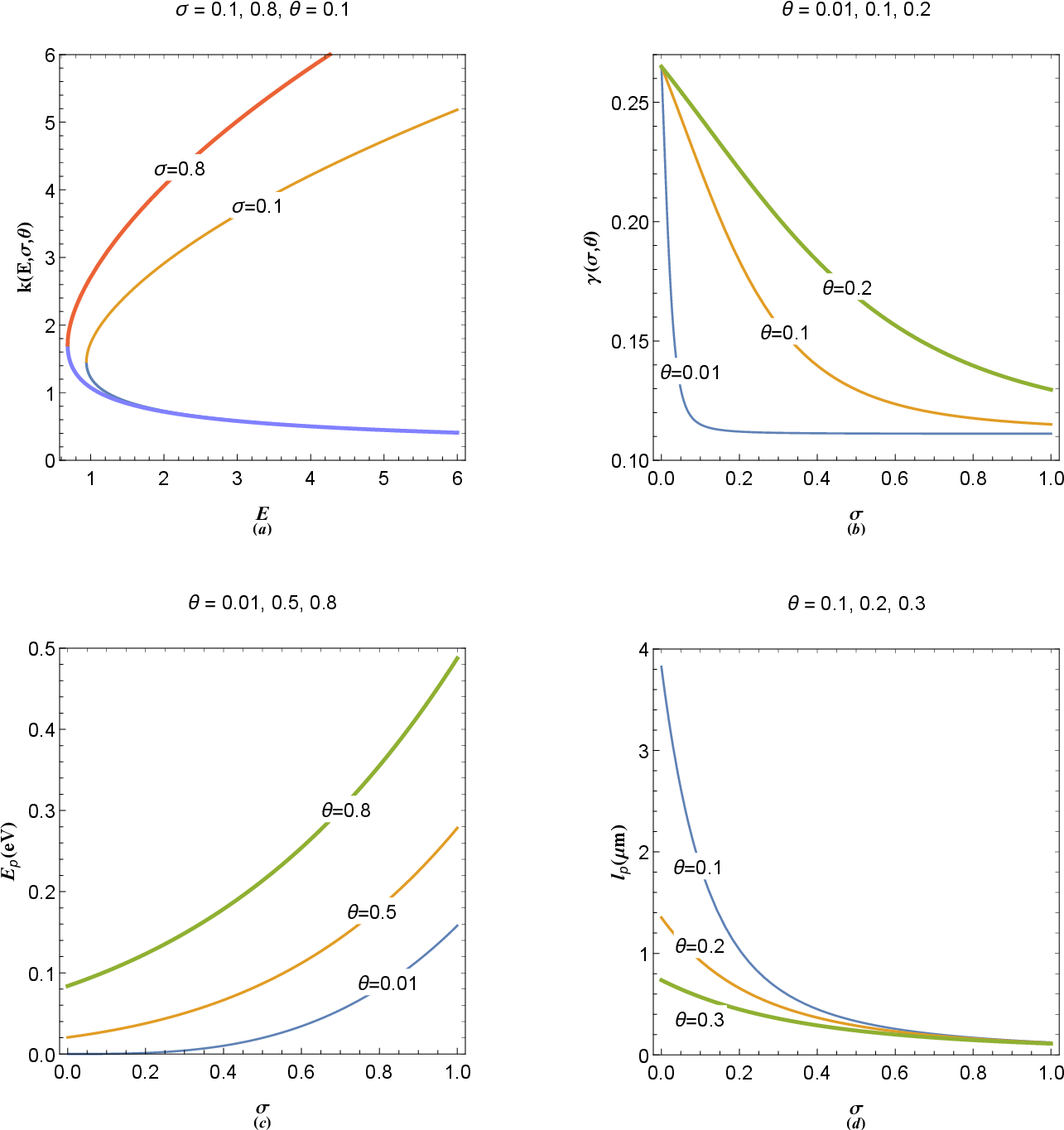}\caption{(a) Effect of normalized chemical potential, $\sigma$, on the characteristic plasmon wavenumbers, for given normalized electron temperature, $\theta$. (b) Effect of normalized chemical potential, $\sigma$ and normalized electron temperature, $\theta$, on the kinetic correction factor, $\gamma$. (c) Variations in the plasmon energy of the electron gas in terms of the normalized chemical potential, $\sigma$, for given normalized electron temperature, $\theta$. (d) Variations in plasmon length of electron gas in terms of the normalized chemical potential, $\sigma$, for given normalized electron temperature, $\theta$. The increase in thickness of curves indicate increase in the varied parameter value above each panel.}
\end{figure}

Figure 1 depicts variations in various parameters involved in plasmon excitations. Figure 1(a) shows the separate excitation branches (wave-like branch with energy $E_1=1/k^2$ and particle-like with energy $E_2=\gamma k^2$), for different values of normalized chemical potential, $\sigma$, and given normalized temperature, $\theta$. Note that the quasiparticle wavenumbers now depend on the chemical potential and electron temperature via the kinetic correction factor, $\gamma(\sigma,\theta)$. The lower/upper branches correspond to wave-/particle-like excitation of the electron gas. It is clearly remarked that chemical potential has significant effect on particle branch, particularly, at higher energy eigenvalues, $E$. The parametric dependence of kinetic correction factor is depicted in Fig. 1(b). It is shown that increase in the normalized electron temperature increases the correction factor, while, increase in the normalized chemical potential leads to decrease in this parameter. Moreover, Fig. 1(c) depicts the dependence of plasmon energy on chemical potential and temperature of the electron gas. It is remarked that, increase in values of both normalized chemical potential and electron temperature leads to increase in the plasmon energy. Note that all these parameter depend explicitly on the number density of the electron gas. The electron number density may be written in normalized form, as
\begin{equation}\label{pd}
n(\sigma ,\theta ) = \frac{{16{e^6}m_e^2{\theta ^6}}}{{{\pi ^3}{\hbar ^6}}}{\rm{L}}{{\rm{i}}_{3/2}}{\left[ {\exp\left( {\frac{{\sigma }}{\theta }} \right)} \right]^4}.
\end{equation}
The variations of plasmon length is depicted in Fig. 1(d), in micrometer units. It is seen that the plasmon length is in the range of few micrometers for classical electron gas  and decreases rapidly to nanometer scale in fully degenerate regime. It is also remarked that it decreases significantly at higher temperature, only, for classical regime.

The solution (\ref{psol}) becomes resonant for electron drift conditions, $k_d=\sqrt{\gamma}k_1$ and $k_d=\sqrt{\gamma}k_2$. It is evident that these resonant wavenumber values depend on the chemical potential and temperature of the electron gas via the kinetic correction factor. However, the dependence on the chemical potential and temperature becomes insignificant for high frequency fast electron phenomenon, where, the kinetic correction approaches unity.

\begin{figure}[ptb]\label{Figure2}
\includegraphics[scale=0.68]{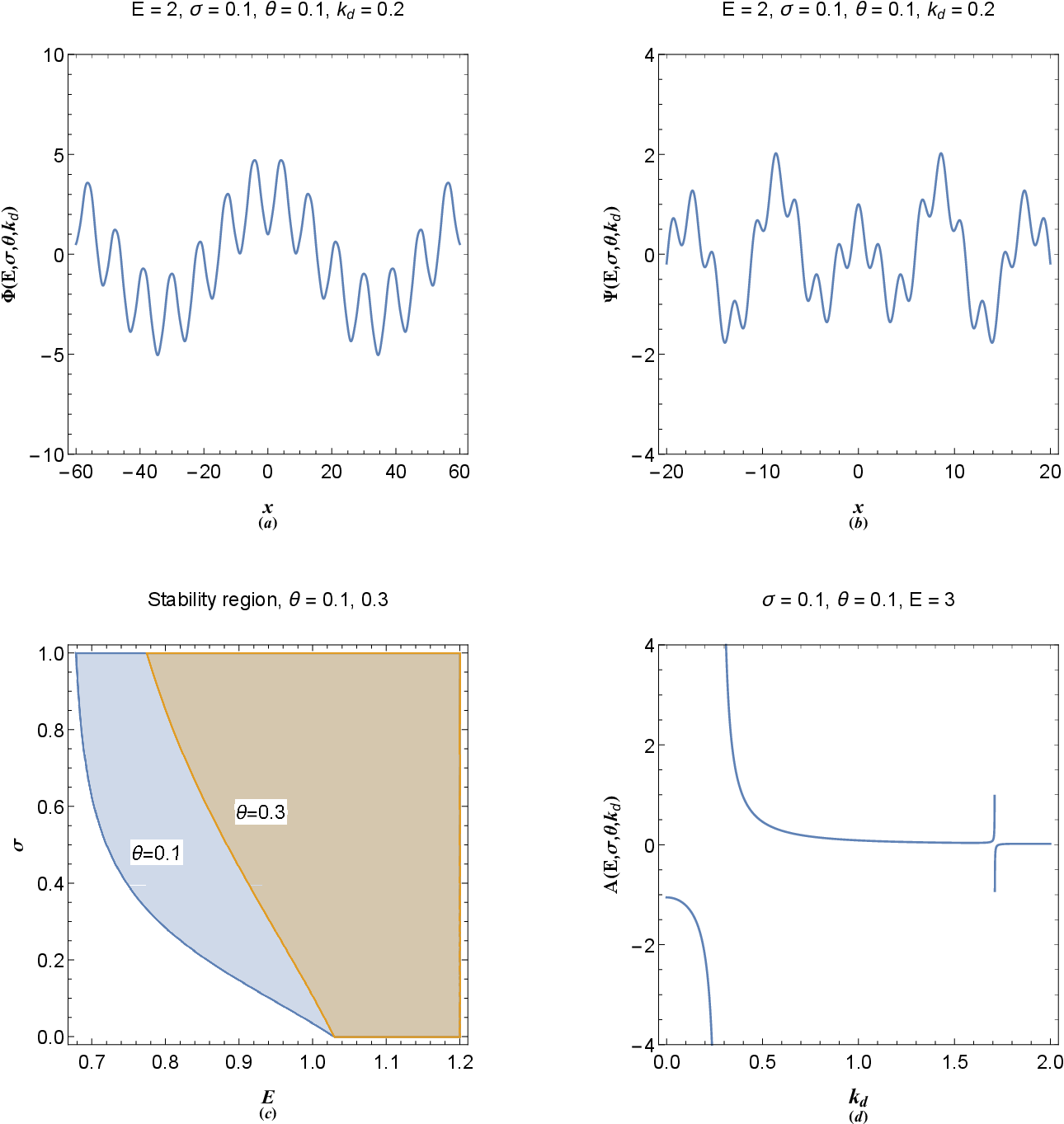}\caption{(a) The electrostatic energy distribution for given energy eigenvalues, chemical potential, electron temperature and drift wavenumber. (b) The statefuntion profiles for given energy eigenvalues, chemical potential, electron temperature and drift wavenumber. (c) Effect of normalized electron temperature on the stabile quasiparticle region in arbitrary degenerate electron gas. (d) Resonant amplitude of driven excitations in the absence of damping effect.}
\end{figure}

Figure 2(a) and 2(b) show the statefunction profiles for the driven plasmon excitations in drifting electron gas with given normalized chemical potential, $\sigma=\mu_0/E_p$, and electron temperature, $\theta=T/T_p$. The oscillations are evidently triple-tone consisting of dual plasmon oscillations in addition to that of the normal driving mode, caused by the electron drift. The drift wavenumber, $k_d$, is in fact the normalized electron drift velocity to that of the plasmon, $v_p=\hbar k_p/m$. Figure 2(c) shows the stability region for collective excitations in the electron gas, as dashed areas, for two different electron temperatures. The excitations are stable for energy eigenvalues, $E>2\sqrt{\gamma}$. It is remarked that the increase in the chemical potential gives rise to the stability of excitations, whereas, the increase in the temperature leads to destabilization effect. Finally, Fig. 2(d) depicts the resonance feature of excitations due to drift electron wavenumber matching that of quasiparticles, i.e. $k_d=\sqrt{\gamma}k_1,\sqrt{\gamma}k_2$. For any given quasiparticle energy eigenvalue, $E$, the wavenumbers, $k_1$ and $k_2$ which satisfy the energy dispersion relation, $E(\sigma,\theta) =k_1^2+k_2^2$, are natural modes of plasmon excitations in the electron gas. In other words, an existing stationary plasmon mode can be excited in the drifting electron gas by the wavenumber matching condition. Note that for $k_d=v/v_p=\gamma^{-1/4}$, drift condition coincides with the collective quantum beating state where, $k_1=k_2$.

\section{Transient and Persistent Excitations}

We now consider the pseudodamped system in which the spacial damping of the wavefunction and electrostatic energy may have various origins. The plasmon damping in electrodynamic concept is based on temporal variations in the amplitude free electron oscillations. The plasmon damping of the later phenomenon in free electron gas has been studied in Ref. \cite{arista}. However, the pseudodamping effect considered in current analysis is due to spatial variations in the electron density, which may be due to polarization effect. One of the main reasons for pseudodamping is the charge screening of positively charged ions or impurities distributed quasineutraly throughout the electron gas in our jellium model. This effect is not involved in the EOS of the gas, since, we have used the Thomas-Fermi assumption which reduces the model to constant chemical potential. While such assumption puts some limitations on applicability of the model, its use for the solid density electron gas, such as for metals and highly-doped semiconductors is confirmed by the well-defined Fermi level \cite{ash}. In our one-dimensional model the pseudodaming is due to column charge polarization and displacement of electron density with respect to the neutralizing background which we treat similar as one dimensional screening. Such 1D screening effect has been studied in the Thomas-Fermi model \cite{kit} with the characteristic normalized wavenumber given as, $\xi(\sigma,\theta)= ({E_p}/2)\partial n/\partial \mu  = (1/2\theta ){\rm{L}}{{\rm{i}}_{1/2}}\left[ { - \exp ({\sigma}/\theta )} \right]/{\rm{L}}{{\rm{i}}_{3/2}}\left[ { - \exp ({\sigma}/\theta )} \right]$. In a pure electron gas, however, the plasmon pseudodamping may also occur due to other mechanisms, such as, electron-electron and electron-phonon collisions. However, appropriate theoretical model of plasmon excitations should include the pseudodamping effect, because, as will be shown the resonance amplitude of plasmon oscillations leads also a resonant current density which in the absence of the pseudodamping term gives rise to nonphysical infinite transport. The model for our purpose reads
\begin{subequations}\label{deq}
\begin{align}
&\gamma \frac{{{d^2}\Psi (x)}}{{d{x^2}}} + 2\gamma \xi \frac{{d\Psi (x)}}{{dx}} + \Phi (x) + E\Psi (x) = {\cal E}\cos \left( {\frac{{i{k_d}x}}{{\sqrt \gamma  }}} \right),\\
&\frac{{{d^2}\Phi (x)}}{{d{x^2}}} + 2\xi\frac{{d\Phi (x)}}{{dx}} - \Psi (x) = 0,
\end{align}
\end{subequations}
in which $\xi$ is the normalized damping parameter. The presence of extra $\gamma$-factor in the pseudodamping term in Eq. (\ref{deq}a) is to guaranty a damped periodic solution analogous to the case of coupled damped harmonic oscillator, for our illustrative purpose. The general solution to homogenous system is
\begin{subequations}\label{gd}
\begin{align}
\begin{array}{l}
{\Phi _{gd}}(x) = \frac{{\gamma {{\rm{e}}^{ - \xi x}}}}{\alpha }\left\{ {\begin{array}{*{20}{l}}
{\left( {k_2^2{\Phi _0} + {\Psi _0}} \right)\left[ {\cos ({\beta _1}x) + \frac{\xi }{{{\beta _1}}}\sin ({\beta _1}x)} \right] - }\\
{\left( {k_1^2{\Phi _0} + {\Psi _0}} \right)\left[ {\cos ({\beta _2}x) + \frac{\xi }{{{\beta _2}}}\sin ({\beta _2}x)} \right]}
\end{array}} \right\},\\
{\Psi _{gd}}(x) = \frac{{\gamma {{\rm{e}}^{ - \xi x}}}}{\alpha }\left\{ {\begin{array}{*{20}{l}}
{\left( {{\Phi _0}/\gamma  + k_2^2{\Psi _0}} \right)\left[ {\cos ({\beta _2}x) + \frac{\xi }{{{\beta _2}}}\sin ({\beta _2}x)} \right] - }\\
{\left( {{\Phi _0}/\gamma  + k_1^2{\Psi _0}} \right)\left[ {\cos ({\beta _1}x) + \frac{\xi }{{{\beta _1}}}\sin ({\beta _1}x)} \right]}
\end{array}} \right\},
\end{array}
\end{align}
\end{subequations}
where $\beta_1=\sqrt{k_1^2-\xi^2}$, $\beta_2=\sqrt{k_2^2-\xi^2}$ are the characteristic damped plasmon wavenumbers. The damped plasmon excitations follow the energy dispersion $E=[1+\gamma(k^2+\xi^2)^2]/2(k^2+\xi^2)$ \cite{akbdual}. It is evident that the general solution (\ref{gd}) is transient and does not extend to large distances. The long expressions for particular solutions, on the other hand, read
\begin{subequations}\label{pd}
\begin{align}
&\Phi (x) =  - \frac{{{\cal E}{{\rm{e}}^{ - \xi x}}}}{{\alpha {\eta _1}{\eta _2}}}\left[ {{\eta _1}\left( {\frac{{k_d^2}}{\gamma } - k_2^2} \right)\cos ({\beta _2}x) - {\eta _2}\left( {\frac{{k_d^2}}{\gamma } - k_1^2} \right)\cos ({\beta _1}x)} \right] + \\
&\frac{{{\cal E}{{\rm{e}}^{ - \xi x}}}}{{\alpha {\eta _1}{\eta _2}}}\left[ {\xi {\eta _1}\left( {\frac{{k_d^2}}{\gamma } + k_2^2} \right)\frac{{\sin ({\beta _2}x)}}{{{\beta _2}}} - \xi {\eta _2}\left( {\frac{{k_d^2}}{\gamma } + k_1^2} \right)\frac{{\sin ({\beta _1}x)}}{{{\beta _1}}}} \right] + \frac{{\cal E}}{{\gamma {\eta _1}{\eta _2}}} \times\\
&\left\{ {\left[ {\left( {\frac{{k_d^2}}{\gamma } - k_1^2} \right)\left( {\frac{{k_d^2}}{\gamma } - k_2^2} \right) - \frac{{4k_d^2{\xi ^2}}}{\gamma }} \right]\cos \left( {\frac{{{k_d}x}}{{\sqrt \gamma  }}} \right) + \frac{{2{k_d}\xi }}{{\sqrt \gamma  }}\left( {k_1^2 + k_2^2 - \frac{{2k_d^2}}{\gamma }} \right)\sin \left( {\frac{{{k_d}x}}{{\sqrt \gamma  }}} \right)} \right\},\\
&\Psi (x) = \frac{{{\cal E}{{\rm{e}}^{ - \xi x}}}}{{\alpha {\eta _1}{\eta _2}}}\left[ {{\eta _1}k_2^2\left( {\frac{{k_d^2}}{\gamma } - k_2^2} \right)\cos ({\beta _2}x) - {\eta _2}k_1^2\left( {\frac{{k_d^2}}{\gamma } - k_1^2} \right)\cos ({\beta _1}x)} \right]-\\
&\frac{{{\cal E}{{\rm{e}}^{ - \xi x}}}}{{\alpha {\eta _1}{\eta _2}}}\left[ {\xi {\eta _1}k_2^2\left( {\frac{{k_d^2}}{\gamma } + k_2^2} \right)\frac{{\sin ({\beta _2}x)}}{{{\beta _2}}} - \xi {\eta _2}k_1^2\left( {\frac{{k_d^2}}{\gamma } + k_1^2} \right)\frac{{\sin ({\beta _1}x)}}{{{\beta _1}}}} \right] - \frac{{\cal E}}{{\gamma {\eta _1}{\eta _2}}} \times\\
&\left\{ {\left[ {1 - \left( {k_1^2 + k_2^2 - \frac{{k_d^2}}{\gamma }} \right)\left( {\frac{{k_d^2}}{\gamma } + 4{\xi ^2}} \right)} \right]\frac{{k_d^2}}{\gamma }\cos \left( {\frac{{{k_d}x}}{{\sqrt \gamma  }}} \right) + \frac{{2\xi k_d^2}}{\gamma }\left( {1 + \frac{{k_d^4}}{{{\gamma ^2}}} - \frac{{4\xi k_d^2}}{\gamma }} \right)\sin \left( {\frac{{{k_d}x}}{{\sqrt \gamma  }}} \right)} \right\},
\end{align}
\end{subequations}
where $\eta_1=(k_d^2/\gamma-k_1^2)^2+4k_d^2\xi^2/\gamma$ and $\eta_2=(k_d^2/\gamma-k_2^2)^2+4k_d^2\xi^2/\gamma$. The persistent solution may be written in the simple forms, $\Phi(x)=A_\phi\cos(k_d x/\sqrt{\gamma}-\phi)$ and $\Psi(x)=A_\psi\cos(k_d x/\sqrt{\gamma}-\psi)$, with amplitudes given by
\begin{subequations}\label{amp}
\begin{align}
&{A_\phi } = \frac{{2\left( {k_d^2 - E} \right)}}{{\gamma {\eta _1}{\eta _2}}}\sqrt {4k_d^2{\xi ^2}{\gamma ^2}{{\left( {\gamma k_1^2 + \gamma k_2^2 - 2k_d^2} \right)}^2} + {{\left[ {k_d^4 - \gamma k_d^2\left( {k_1^2 + k_2^2 + 4{\xi ^2}} \right) + {\gamma ^2}k_1^2k_2^2} \right]}^2}},\\
&{A_\psi } = \frac{{k_d^2\left( {E - k_d^2} \right)}}{{{\gamma ^4}{\eta _1}{\eta _2}}}\sqrt {\frac{{4\gamma {\xi ^2}}}{{k_d^2}}{{\left( {k_d^4 + 4\gamma {\xi ^2}k_d^2 - {\gamma ^2}} \right)}^2} + {{\left\{ {{\gamma ^2} + \left[ {k_d^2 - \gamma \left( {k_1^2 + k_2^2} \right)} \right]\left( {k_d^2 + 4\gamma {\xi ^2}} \right)} \right\}}^2}},
\end{align}
\end{subequations}
and the corresponding phase values as follows
\begin{subequations}\label{phi}
\begin{align}
&\tan\phi  =\left[ {\frac{{2\xi (k_d/\sqrt{\gamma})\left( {k_1^2 + k_2^2 - 2{k_d^2/\gamma}} \right)}}{{\left( {{k_d^2}/\gamma - k_1^2} \right)\left( {{k_d^2}/\gamma - k_2^2} \right) - 4{k_d^2}{\xi^2}/\gamma}}} \right],\\
&\tan\psi  =\left[ {\frac{{2(\sqrt{\gamma}\xi/k_d)\left( {1 + {k_d^4}/\gamma^2 - 4{\xi^2}{k_d^2}/\gamma} \right)}}{{1 - \left( {k_1^2 + k_2^2 - {k_d^2}/\gamma} \right)\left( {{k_d^2}/\gamma + 4{\xi^2}} \right)}}} \right].
\end{align}
\end{subequations}
Note that the excitation amplitude and phases also depend on the chemical potential and the temperature of the electron gas via the kinetic correction and damping parameter. In a driven damped harmonic coupled oscillator the resonance occurs at natural frequencies of the system. However, in our case these are the natural quasiparticle wavenumbers of the system, $k_1$ and $k_2$ at which the resonance takes place.

\section{Current Density of Resonant Orbital}

In previous sections we considered coupled pseudoforce system with real driver term in order to bring to view the similarities with coupled harmonic oscillators. Let us now consider the solution to the original system with the imaginary pseudoforce
\begin{subequations}\label{ex}
\begin{align}
&\gamma \frac{{{d^2}\Psi (x)}}{{d{x^2}}} + 2\gamma \xi \frac{{d\Psi (x)}}{{dx}} + \Phi (x) + E\Psi (x) = {\cal E}\exp \left( {\frac{{i{k_d}x}}{{\sqrt \gamma  }}} \right),\\
&\frac{{{d^2}\Phi (x)}}{{d{x^2}}} + 2\xi\frac{{d\Phi (x)}}{{dx}} - \Psi (x) = 0.
\end{align}
\end{subequations}
The persistent solutions are assumed to be of the forms, $\Phi(x)=A\exp(ik_d x/\sqrt{\gamma})$ and $\Psi(x)=B\exp(ik_d x/\sqrt{\gamma})$. Plugging these solutions into (\ref{ex}), we arrive at
\begin{subequations}\label{ex1}
\begin{align}
&\gamma A  + \gamma E \left( {1 + B} \right) - {k_d}\left( {{k_d}\gamma  +\gamma{k_d} B  - 2{\rm{i}}\xi{\gamma ^{3/2}} B } \right) = 0,\\
&\gamma B  +{k_d}\left( {{k_d} - 2{\rm{i}} \xi \sqrt \gamma } \right) A = 0.
\end{align}
\end{subequations}
By solving the system (\ref{ex1}), we easily obtain the real and imaginary components of amplitudes, $A=A_r+iA_i$ and $B=B_r+iB_i$, as follows
\begin{subequations}\label{ex1}
\begin{align}
&{A_r} = \frac{{{\gamma ^2}\left( {E - k_d^2} \right)\left( {\gamma Ek_d^2 - {\gamma ^2} - \gamma k_d^4 + 4{\gamma ^2}{\xi ^2}k_d^2} \right)}}{{{{\left( {\gamma k_d^4 - \gamma Ek_d^2 + {\gamma ^2}} \right)}^2} + 4{\gamma ^3}{\xi ^2}k_d^2\left[ {{E^2} - 2\left( {Ek_d^2 + \gamma } \right) + 2k_d^4} \right] + 16{\gamma ^4}{\xi ^4}k_d^4}},\\
&{A_i} = \frac{{2{\gamma ^{7/2}}\xi {k_d}\left( {E - k_d^2} \right)^2}}{{{{\left( {\gamma k_d^4 - \gamma Ek_d^2 + {\gamma ^2}} \right)}^2} + 4{\gamma ^3}{\xi ^2}k_d^2\left[ {{E^2} - 2\left( {Ek_d^2 + \gamma } \right) + 2k_d^4} \right] + 16{\gamma ^4}{\xi ^4}k_d^4}},\\
&{B_r} = \frac{{\gamma k_d^2\left( {E - k_d^2} \right)\left[ {{\gamma ^2} + \gamma k_d^4 + 4{\gamma ^2}{\xi ^2}k_d^2 - \gamma E\left( {k_d^2 + 4\gamma {\xi ^2}} \right)} \right]}}{{{{\left( {\gamma k_d^4 - \gamma Ek_d^2 + {\gamma ^2}} \right)}^2} + 4{\gamma ^3}{\xi ^2}k_d^2\left[ {{E^2} - 2\left( {Ek_d^2 + \gamma } \right) + 2k_d^4} \right] + 16{\gamma ^4}{\xi ^4}k_d^4}},\\
&{B_i} = \frac{{2\xi {\gamma ^{5/2}}{k_d}\left( {E - k_d^2} \right)\left( {k_d^4 + 4\gamma {\xi ^2}k_d^2 - \gamma } \right)}}{{{{\left( {\gamma k_d^4 - \gamma Ek_d^2 + {\gamma ^2}} \right)}^2} + 4{\gamma ^3}{\xi ^2}k_d^2\left[ {{E^2} - 2\left( {Ek_d^2 + \gamma } \right) + 2k_d^4} \right] + 16{\gamma ^4}{\xi ^4}k_d^4}},
\end{align}
\end{subequations}
where the phase differences with the driver are, $\phi=\arctan(A_i/A_r)$ and $\psi=\arctan(B_i/B_r)$. The normalized current density may be calculated as, $j=(i/2)[\Psi(x)d\Psi^*(x)/dx-\Psi^*(x)d\Psi(x)/dx]=k_d|B|^2/\sqrt{\gamma}$. Note that the current density of quasiparticle orbital does not spatially vary in resonant state and therefore there is no local charge variation over time (stationary state). The thermodynamic equilibrium condition is characterized by mixed quasiparticle state by numeration of density of states over all quasiparticle orbital
\begin{equation}\label{tj}
J = \int\limits_{2\sqrt \gamma  }^\infty  {\frac{{{k_d}}}{{\sqrt \gamma  }}{|B(E)|^2}D(E)f(E)dE},
\end{equation}
where $f(E)=1/[1+\exp(E/\theta)]$ is the fermion occupation number (for the degenerate gas) and $D(E)$ is the quasiparticle density of states (DoS), given by \cite{pub}
\begin{equation}\label{dos}
D(E) = \frac{{\left( {{E^2} - 4\gamma } \right)\left( {{L^4} + \gamma } \right) + E\left( {{L^4} - \gamma } \right)\sqrt {{E^2} - 4\gamma } }}{{4\pi L\sqrt 2 \left( {{E^2} - 4\gamma } \right)\sqrt {\gamma \left( {{L^4} - \gamma } \right)\sqrt {{E^2} - 4\gamma }  - 4{L^2}{\gamma ^2} + E\left( {{L^4} + \gamma } \right)} }}.
\end{equation}
in which $L$ is the normalized size of the one dimensional system and the lower integrations limit ($E_m=2\sqrt{\gamma}$) refers to the bottom of quasiparticle conduction band.

Current model may be readily extended to multistreaming electron gas as follows. Assuming that $\Psi_j$ is the pure wavefunction of the $j$-th stream, one may start from the following normalized N-pure wavefunction damped pseudoforce model
\begin{subequations}\label{ms}
\begin{align}
&\left[ {\gamma \frac{{{d^2}}}{{d{x^2}}} + 2\gamma \xi \frac{d}{{dx}} + \Phi (x) + E} \right]{\Psi _j}(x){e^{\frac{{i{p_j}x}}{{\sqrt \gamma  }}}} = 0,\\
&\frac{{{d^2}\Phi (x)}}{{d{x^2}}} + 2\xi \frac{{d\Phi (x)}}{{dx}} = \sum\limits_j {{F_j}|{\Psi _j}(x)|{^2}}  - 1,
\end{align}
\end{subequations}
in which $n(x)=\sum\limits_j {{F_j}|{\Psi _j}(x)|{^2}}$ with $F_j$ being the probability of $j$-th momentum orbital, satisfying the condition $\sum\nolimits_j F_j=1$. Also, $p_{j}$ is the $j$-th stream momentum as normalized to characteristic plasmon momentum, $p_p=\hbar k_p$. After appropriate linearization and weighted summation of \ref{ms}(a) over $j$-th momentum orbital, the system reads
\begin{subequations}\label{msn}
\begin{align}
&\gamma \frac{{{d^2}\Xi (x)}}{{d{x^2}}} + 2\gamma \xi \frac{{d\Xi (x)}}{{dx}} + \Phi (x) + E\Xi (x) = \sum\nolimits_j F_j{{{\cal E}_j}\exp \left( {\frac{{i{p_{j}}x}}{{\sqrt \gamma  }}} \right)},\\
&\frac{{{d^2}\Phi (x)}}{{d{x^2}}} + 2\xi\frac{{d\Phi (x)}}{{dx}} - \Xi (x) = 0,
\end{align}
\end{subequations}
where ${\cal E}_j=(E-p_{j})$ and $\Xi (x) = \sum\nolimits_j {{F_j}{\Psi _j}(x)}$. The particular stable solutions to (\ref{msn}) are of the forms, $\Phi (x) = \sum\limits_j F_j{{A_j}\exp (i{p_{j}}x/\sqrt \gamma  )}$ and $\Xi (x) = \sum\limits_j F_j{{B_j}\exp (i{p_{j}}x/\sqrt \gamma  )}$, where, $A_j=A_{rj}+i A_{ij}$ and $B_j=B_{rj}+i B_{ij}$ with $A_{rj}$($B_{rj}$) and $A_{ij}$($B_{ij}$) being the real and imaginary components, respectively,
\begin{subequations}\label{ex1}
\begin{align}
&{A_{rj}} = \frac{{{\gamma ^2}\left( {E - p_{j}^2} \right)\left( {\gamma E p_{j}^2 - {\gamma ^2} - \gamma p_{j}^4 + 4{\gamma ^2}{\xi ^2}p_{j}^2} \right)}}{{{{\left( {\gamma p_{j}^4 - \gamma E p_{j}^2 + {\gamma ^2}} \right)}^2} + 4{\gamma ^3}{\xi ^2}p_{j}^2\left[ {{E^2} - 2\left( {E p_{j}^2 + \gamma } \right) + 2p_{j}^4} \right] + 16{\gamma ^4}{\xi ^4}p_{j}^4}},\\
&{A_{ij}} = \frac{{2{\gamma ^{7/2}}\xi {p_{j}}\left( {E - p_{j}^2} \right)^2}}{{{{\left( {\gamma p_{j}^4 - \gamma E p_{j}^2 + {\gamma ^2}} \right)}^2} + 4{\gamma ^3}{\xi ^2}p_{j}^2\left[ {{E^2} - 2\left( {E p_{j}^2 + \gamma } \right) + 2p_{j}^4} \right] + 16{\gamma ^4}{\xi ^4}p_{j}^4}},\\
&{B_{rj}} = \frac{{\gamma p_{j}^2\left( {E - p_{j}^2} \right)\left[ {{\gamma ^2} + \gamma p_{j}^4 + 4{\gamma ^2}{\xi ^2}p_{j}^2 - \gamma E\left( {p_{j}^2 + 4\gamma {\xi ^2}} \right)} \right]}}{{{{\left( {\gamma p_{j}^4 - \gamma E p_{j}^2 + {\gamma ^2}} \right)}^2} + 4{\gamma ^3}{\xi ^2}p_{j}^2\left[ {{E^2} - 2\left( {E p_{j}^2 + \gamma } \right) + 2p_{j}^4} \right] + 16{\gamma ^4}{\xi ^4}p_{j}^4}},\\
&{B_{ij}} = \frac{{2\xi {\gamma ^{5/2}}{p_{j}}\left( {E - p_{j}^2} \right)\left( {p_{j}^4 + 4\gamma {\xi ^2}p_{j}^2 - \gamma } \right)}}{{{{\left( {\gamma p_{j}^4 - \gamma E p_{j}^2 + {\gamma ^2}} \right)}^2} + 4{\gamma ^3}{\xi ^2}p_{j}^2\left[ {{E^2} - 2\left( {E p_{j}^2 + \gamma } \right) + 2p_{j}^4} \right] + 16{\gamma ^4}{\xi ^4}p_{j}^4}}.
\end{align}
\end{subequations}
The total current density of mixed momentum-state drifting electron gas in the large system size limit, $L\gg 1$, with electron momentum distribution, $F(p)$, is given by
\begin{equation}\label{tj}
{J_T} = \frac{1}{{\sqrt \gamma  }}\int\limits_0^\infty  {F(p)pdp} \int\limits_{2\sqrt \gamma  }^\infty  {|B(E,p)|^2D(E)f(E)dE},
\end{equation}
where $\int {F(x,p)dp = \rho(x)}$ with $\rho(x)$ being the mass density distribution. The total current-density equation, (\ref{tj}), is general in the sense that it also takes into account the electron-plasmon resonant interactions.

\begin{figure}[ptb]\label{Figure3}
\includegraphics[scale=0.68]{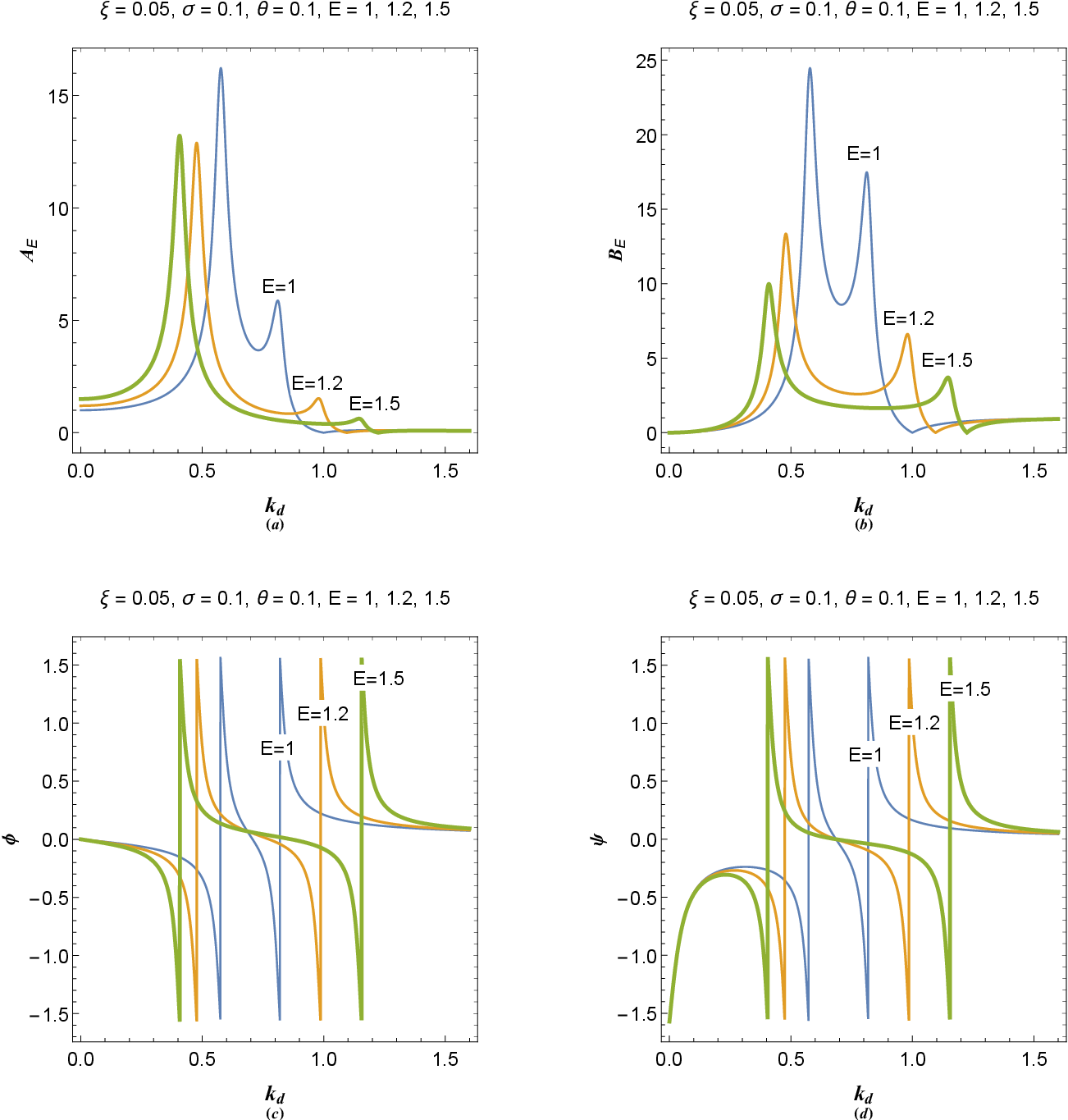}\caption{(a) Resonant amplitude of driven electron density excitation in terms of normalized electron drift wavenumber for different energy eigenvalues and other fixed parameters. (b) Resonant amplitude of driven electrostatic field excitations in terms of normalized electron drift speed for different energy eigenvalues and other fixed parameters. (c) The resonant density excitation phase corresponding to parameter values used in (a). (d) The resonant electrostatic field excitation phase corresponding to parameter values used in (b). The increase in thickness of curves indicate increase in the varied parameter value above each panel.}
\end{figure}

Figure 3 shows resonant amplitudes and phases of damped quasiparticle excitations (\ref{pd}) for various parameter values in terms of normalized electron drift wavenumber. Figures 3(a) and 3(b) reveal that two set of resonant amplitude peaks occur for given parameters at quasiparticle energy eigenvalues corresponding to wave-like and particle-like excitations at points $k_d=\sqrt{\gamma}k_1$ and $k_d=\sqrt{\gamma}k_2$, respectively. It is also remarked that the wave-like resonance peaks are relatively higher. The long wavelength peak are caused by drift resonance condition with the wave-like branch of dispersion curve, while, the small wavelength ones are due to drift wavenumber matching condition at particle-like branch. It is revealed that increase in the energy eigenvalue of excitations leads the height of peaks to decrease and move to further apart from each other. An interesting feature is the Fano-like asymmetric shape of the resonance peaks, particularly, that of the particle-like excitations. The Fano resonance is caused by interference of scattered waves from multiple sites or presence of dual resonance effect. Here, the effect refers to the electron-plasmon resonant interactions in the presence of background wave-like scattering phenomenon. Figure 3(c) and 3(d) show the phase differences with driver corresponding to amplitudes in Figs. 3(a) and 3(b) with the same parameters, respectively. The sharp discontinuous change in the phases corresponding to resonant amplitudes, are clearly apparent. It is seen that electron density and electrostatic energy of driven electron gas excitations are in phase with each other for all drift wavenumber range except in the very low values in which the electrostatic field phase vanishes, whereas, that of the density maximizes. The later case occurs for electron drift velocity much smaller compared to the plasmon speed, $v\ll \sqrt{2E_p/m}$. It is clearly remarked that very close to resonances, the driver pseudoforce, $\Psi_d(x)={\cal E}\exp(ik_d x)$, and the pseudovelocity, $d\Psi(x)/dx$ are completely out of phase, i.e. $\psi=\pm \pi/2$. However, at exact resonance point this phase shift vanishes. Therefore, the drift current density in present model plays the exact analogous role of the external force power in driven damped harmonic oscillators.

\begin{figure}[ptb]\label{Figure4}
\includegraphics[scale=0.68]{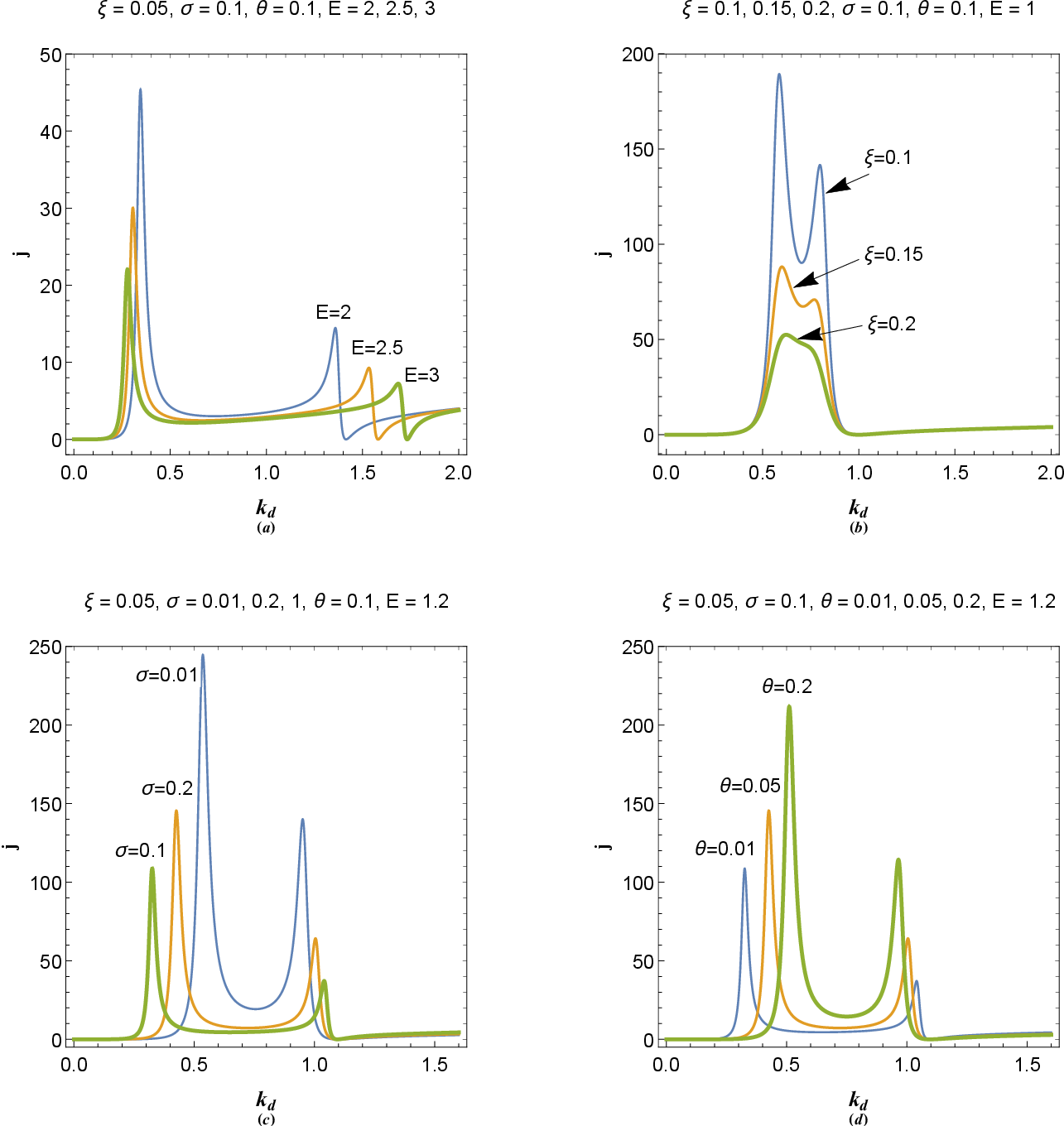}\caption{(a) Variation of the resonant current density in terms of drift wavenumber for different of energy eigenvalue and fixed chemical potential, temperature and damping parameter. (b) Variation of the resonant current density in terms of drift wavenumber for different of damping parameter and fixed chemical potential, temperature and energy eigenvalue. (c) Variation of the resonant current density in terms of drift wavenumber for different of normalized chemical potentoal and fixed energy eigenvalue, electron temperature and damping parameter. (d) Variation of the resonant current density in terms of drift wavenumber for different of electron temperature and fixed chemical potential, excitation energy and damping parameter. The increase in thickness of curves indicate increase in the varied parameter value above each panel.}
\end{figure}

Figure 4 shows the variations in resonant current density in terms of drift wavenumber for different values of parameters. Similar to the resonant amplitudes shown in Fig. 3, the current density shows pronounced peaks at resonant drift wavenumbers. Also, the asymmetric shape of peaks refer to multiple scattering interferences. It is seen from Fig. 4(a) that increase in energy eigenvalue decreases the peak heights shifting the relative wavenumber values to lower/higher values for wave/particle resonance peaks. It generally concluded that resonance condition becomes intense as the two peaks are closer to each other. That is to say the resonance maximizes at quantum beating point, $k_1=k_2$ or $k_d=\gamma^{1/4}$. Figure 4(b) shows the effect of damping parameter on the resonant current density. It shows that increase in damping parameter strongly lowers the amplitude of resonant peaks, where, they are merged for very large damping parameter values. Figures 4(c) and 4(d) show similar features for current density as shown by Figs. 3(c) and 3(d) for resonance amplitudes. However, in this case the peaks are clear and sharper. The sharpness of resonance peaks is an indication of the resonant interaction quality and is a measure of electron drift wavenumber (momentum) spread.

\begin{figure}[ptb]\label{Figure5}
\includegraphics[scale=0.67]{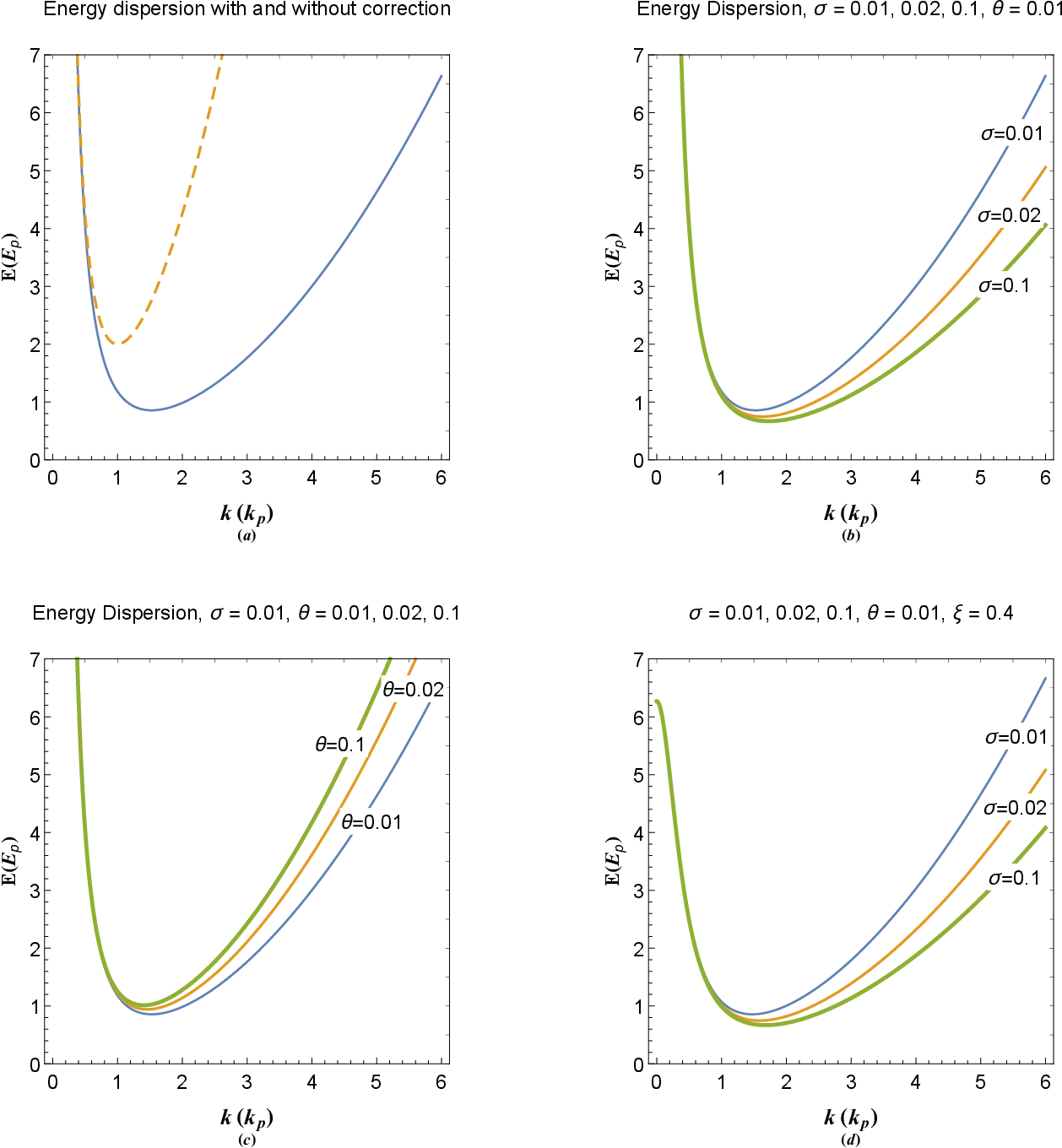}\caption{(a) The energy dispersion in the absence (dashed curve) and presence (solid curve) of the kinetic correction. (b) The effect of normalized electron chemical potential on the plasmon dispersion curve. (c) The effect of normalized electron temperature on the plasmon dispersion curve. (d) The effect of normalized electron chemical potential on the damped plasmon dispersion curve. Increase in thickness of curves indicate increase in varied parameter above each panel.}
\end{figure}

In Figure 5 we investigate the effects of the kinetic correction on the energy dispersion of quasiparticle excitations. Figure 5(a) depicts the matter wave dispersion curve in the presence (solid curve) and the absence (dashed curve) of kinetic correction for given chemical potential and electron temperature parameter values. The effect of kinetic correction is huge on the low phase-speed particle-like branch (as we call it the kinetic regime) with almost no effect on the high phase-speed wave-like branch, which are interconnected at the quantum beating point, i.e. minimum of quasiparticle conduction band at $k_m=\gamma^{-1/4}$. The effect of kinetic correction is evidently becomes more pronounced as the energy eigenvalue increases. The group speed of low phase-speed branch has significantly decreased due to this correction. The feature has been previously pointed out in hydrodynamic analysis of one component quantum plasmas \cite{akbhd,haas2016}, where it was pointed out that the kinetic correction only applies to low phase-speed plasmon phenomenon, in the static limit. It is remarkable that the kinetic correction profoundly lowers the energy minimum (plasmon conduction levels) of excitations orbital accessible to quasiparticles. Figure 5(b) reveals the effect of normalized chemical potential of electron gas on the energy dispersion indicating that increase of the parameter leads to decrease of group-speed of low phase speed plasmon excitations. Note that the second derivative of energy is a measure of effective electron mass and their mobility at a given wavenumber in plasmon conduction level. It is therefore concluded that increase in the chemical potential leads to increase in the effective mass of electrons and decrease in their mobility. Moreover, Fig. 5(c) shows the effect of electron temperature on the dispersion curve. It is evident that increase in the electron gas temperature increases the quasiparticle mobility in contrast to the case of chemical potential increase. Finally, Fig. 5(d) shows the effect of chemical potential on dispersion curve of damped plasmons. The similar effect as in Fig. 5(b) is also apparent in this case. However, the presence of strong damping (charge screening) leads to truncation of wave-like branch.

\begin{figure}[ptb]\label{Figure6}
\includegraphics[scale=0.68]{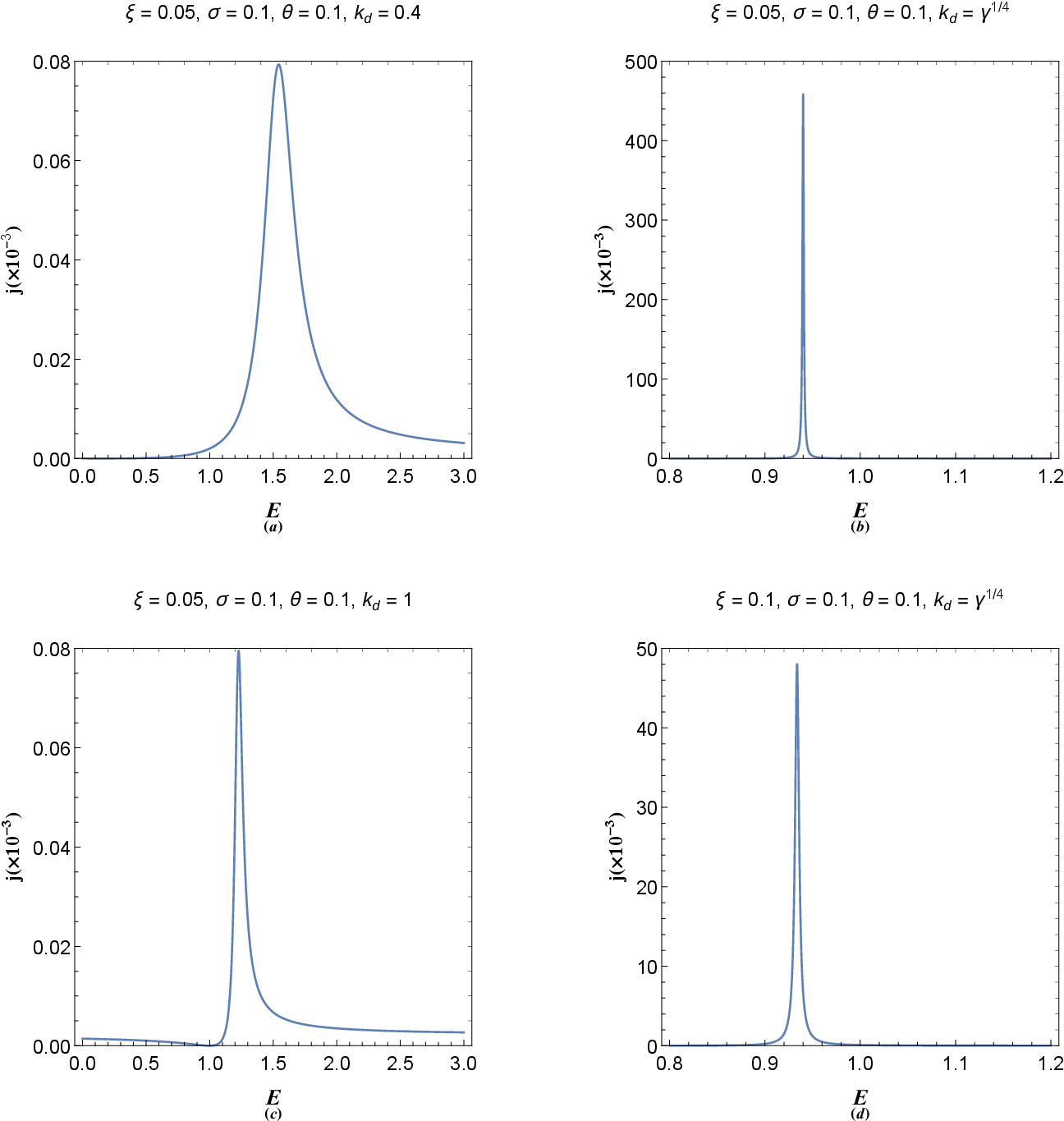}\caption{Resonance corresponding to the current density of drifting arbitrary degenerate electron gas in terms of quasiparticle energy eigenvalues for fixed values of other parameters, such as, the chemical potential, the electron temperature, drift wavenumber and damping parameter. (a) Drift wavenumber smaller than plasmon wavenumber. (b) Drift wavenumber equal to plasmon wavenumber. (c) Drift wavenumber larger than plasmon wavenumber. (d) Drift wavenumber equal to plasmon wavenumber and higher damping parameter.}
\end{figure}

Figure 6 shows the energy spectrum of current density resonance in interaction between drifting electrons and quasiparticles It is seen that the energy spectrum of quasiparticle driven by electron drift contains well defined current density peaks for given parameter. Figure 6(a) shows the resonant current peak for drift wavenumber matching at wave-like branch, $k_d=0.4$. It contains a broad asymmetric peak around $E=1.5$. The resonance quality factor may be defined as $Q=E_r/\Delta E$, where, $E_r$ is the resonant quasiparticle energy and $\Delta E$ being the half-current bandwidth. This factor indicates how sharply defined is the drift wavenumber corresponding to a single-momentum electron beam. Figure 6(b) shows the resonant peak for $k_d=\gamma^{1/4}\simeq 0.69$, corresponding to dispersion minimum wavenumber (quantum beating point). It is remarkable that at this point the resonance profile contains a very localized sharp maximum with a very high quality factor. It is evident that the corresponding drift wavenumber of electron gas moving at speed very close to the plasmon speed, is sharply defined. Moreover, Fig. 6(c) reveals that the for drift wavenumber resonance at particle-like branch $k_d=1$ the peak height/width decreases/increases. However, the quality of resonances for electron drifts larger than plasmon speed are relatively better than those with smaller speed than that of plasmon. Finally, Fig. 6(d) shows the effect of damping on the resonance at quantum beating condition in Fig. 6(b). It is seen that increase in damping leads to sharp decrease in peak height and increase in it width. It is also remarked that the resonance peak slightly moves to lower wavenumbers by increase of the damping parameter. As a final conclusion, one may say that drift in electron gas leads to a resonant collective excitation in the gas at quasiparticle energy corresponding to drift wavenumber.

\section{Conclusion}

We studied the resonant interaction between drifting electrons and collective excitations in an electron gas with arbitrary degeneracy, using the effective kinetic-corrected Schr\"{o}dinger-Poisson system. It was found that the kinetic correction leads to significant effect of electron number density and temperature on the collective excitations in the low phase speed regime. We derived the solution of the damped driven pseudoforce system in the presence of kinetic correction and calculated the amplitude and phase shifts of electrostatic energy and  density function relative to that of the driver pseudoforce. The kinetic correction was found to profoundly affect the dispersion curve in the low phase-speed (small wavelength) regime as we call it the kinetic regime. It was further revealed that electron density leads to increase in effective mass of quasiparticles and decrease in their mobility in the plasmon conduction band, while, the increase in the electron temperature has the inverse effect. Current mode of drifting quantum electron gas can be used to investigate various other phenomena such as the beam-plasma interactions.

\section{Data Availability}

The data that support the findings of this study are available from the corresponding author upon reasonable request.

\end{document}